\documentclass[aps,prl,twocolumn,superscriptaddress,longbibliography,amsmath,amssymb]{revtex4-2}
\usepackage{xcolor}
\usepackage{graphicx}

\usepackage[separate-uncertainty = true,multi-part-units=single]{siunitx}
\usepackage[version=4,arrows=pgf-filled,
textfontname=sffamily,
mathfontname=mathsf]{mhchem}

\usepackage[utf8]{inputenc}
\usepackage{amssymb,amsmath}
\usepackage{tabularx}
\usepackage{multirow}
\usepackage{booktabs}
\usepackage{physics}
\usepackage{colortbl}
\usepackage{tabularx}
\usepackage{color}
\usepackage{systeme}
\usepackage{cleveref}

\usepackage{multibib}
\newcites{A}{References}
\newcites{B}{References}

\begin{document}
\title{High luminescence efficiency of multi-valley excitonic complexes in heavily doped WSe$_2$ monolayer}

\author{S\'ebastien Roux}
\thanks{Equal contribution}
\affiliation{Universit\'e de Toulouse, INSA-CNRS-UPS, LPCNO, 135 Av. Rangueil, 31077 Toulouse, France}
\author{Tilly Guyot}
\thanks{Equal contribution}
\affiliation{Universit\'e de Toulouse, INSA-CNRS-UPS, LPCNO, 135 Av. Rangueil, 31077 Toulouse, France}
\author{Abraao Cefas Torres-Dias}
\affiliation{Universit\'e de Toulouse, INSA-CNRS-UPS, LPCNO, 135 Av. Rangueil, 31077 Toulouse, France}
\author{Delphine Lagarde}
\affiliation{Universit\'e de Toulouse, INSA-CNRS-UPS, LPCNO, 135 Av. Rangueil, 31077 Toulouse, France}
\author{Laurent Lombez}
\affiliation{Universit\'e de Toulouse, INSA-CNRS-UPS, LPCNO, 135 Av. Rangueil, 31077 Toulouse, France}
\author{Dinh Van Tuan}
\affiliation{Department of Electrical and Computer Engineering,
University of Rochester, Rochester, NY, United States}
\author{Junghwan Kim}
\affiliation{Department of Electrical and Computer Engineering,
University of Rochester, Rochester, NY, United States}
\author{Kenji Watanabe}
\affiliation{Research Center for Electronic and Optical Materials, National Institute for Materials Science, 1-1 Namiki, Tsukuba 305-0044, Japan}
\author{Xavier Marie}
\affiliation{Universit\'e de Toulouse, INSA-CNRS-UPS, LPCNO, 135 Av. Rangueil, 31077 Toulouse, France}
\affiliation{Institut Universitaire de France, 75231, Paris, France}
\author{Takashi Taniguchi}
\affiliation{Research Center for Materials Nanoarchitectonics, National Institute for Materials Science, 1-1 Namiki, Tsukuba 305-0044, Japan}
\author{Hanan Dery}
\affiliation{Department of Electrical and Computer Engineering,
University of Rochester, Rochester, NY, United States}
\affiliation{Department of Physics and Astronomy, University of Rochester, Rochester, NY, United States}
\author{Cedric Robert}
\thanks{Corresponding author: cerobert@insa-toulouse.fr}
\affiliation{Universit\'e de Toulouse, INSA-CNRS-UPS, LPCNO, 135 Av. Rangueil, 31077 Toulouse, France}

%


\begin{abstract}

Monolayers of group-VI transition-metal dichalcogenides (TMDs) are two-dimensional semiconductors that exhibit exceptionally strong light--matter coupling yet typically suffer from low emission quantum yields. In this letter, we investigate the heavily $n$-doped regime of a WSe$_2$ monolayer and show that multi-particle excitonic complexes produce photoluminescence signals up to two orders of magnitude stronger than in the neutral state. Time-resolved photoluminescence and differential reflectivity measurements reveal that the quantum yield rises with carrier density and exceeds 50\% for electron concentrations above $10^{13}\,\mathrm{cm}^{-2}$. These findings establish TMD monolayers as a platform for exploring excitonic complexes in high-density electron gases and point toward new opportunities for efficient, atomically thin light emitters.

\end{abstract}

\maketitle 

Atomically thin layers of group VI transition metal dichalcogenides (TMDs) such as MoS$_2$, WSe$_2$, MoSe$_2$, WS$_2$, and MoTe$_2$ are two-dimensional (2D) semiconductors that have attracted tremendous interest over the past 15 years owing to their exceptional optical properties, characterized by an unusually strong light–matter interaction \citeA{Mak2010,splendiani_emerging_2010,wang_colloquium_2018}. This strong interaction originates from the large Coulomb forces due to the reduced dimensionality and dielectric screening, which give rise to tightly bound electron-hole pairs (excitons) with binding energies of several hundreds of meV \citeA{goryca_revealing_2019, chernikov_exciton_2014,qiu_optical_2013}.
In addition, the carrier density (electrons or holes) in these monolayers (MLs) can be easily tuned electrostatically by embedding the material in a charge-tunable device. The strong Coulomb interaction also drives the formation of a variety of excitonic complexes involving both excitons and free carriers, such as trions \citeA{mak_tightly_2013, ross_electrical_2013, courtade_charged_2017,perea-causin_trion_2024}. Moreover, the band structure of TMD MLs is governed by a multivalley configuration combined with large spin–orbit coupling, leading to a diverse landscape of excitonic species—neutral or charged, optically bright, spin-forbidden dark, or momentum-indirect—opening the door to a uniquely rich excitonic physics \citeA{liu_gate_2019, li_direct_2019, yang_relaxation_2022, malic_dark_2018, ye_efficient_2018, molas_brightening_2017}.

Despite their strong light–matter coupling, TMD MLs are notoriously limited by a relatively low photoluminescence (PL) quantum yield (\textit{i.e.} emitted photons per absorbed photons) compared to conventional III–V quantum wells. This reduction is generally attributed to three main factors: (i) a high density of intrinsic point defects \citeA{liu_two-step_2023}, (ii) efficient density dependent exciton–exciton annihilation processes \citeA{goodman_substrate-dependent_2020}, and (iii) the presence of optically dark states. Several approaches have been proposed to enhance the luminescence quantum yield. Amani \textit{et al.} demonstrated that chemical passivation of defects can enhance the quantum yield \citeA{amani_near-unity_2015, tanoh_giant_2021}, whereas Lien \textit{et al.} reported that near-unity quantum yield can also be achieved by compensating intrinsic doping through electrostatic gating \citeA{lien_electrical_2019}. Nevertheless, most estimations of the absolute quantum yield rely on non-trivial calibration of PL intensities, typically by comparison with dye emission standards, which introduces substantial uncertainties.

In this letter, we investigate radically different doping densities: the almost unexplored regime of heavy electron doping in monolayer WSe$_2$ (carrier densities above 10$^{13}$ cm$^{-2}$). In this highly doped regime, excitonic complexes involving six, eight, or even more particles have recently been identified \citeA{van_tuan_six-body_2022, choi_emergence_2024, dijkstra_ten-valley_2025}. We demonstrate that the PL intensity associated with the recombination of these many-body complexes increases with electron density. Furthermore, by combining reflectivity measurements with time-resolved photoluminescence (TRPL), we estimate that the quantum yield exceeds 50\% at the highest doping levels, a result attributed to the suppression of relaxation into dark excitonic states. These findings demonstrate that heavy electron doping can substantially enhance the quantum yield of TMD monolayers, even in the absence of defect passivation treatment.

\begin{figure}[h!]
\includegraphics[width=\linewidth]{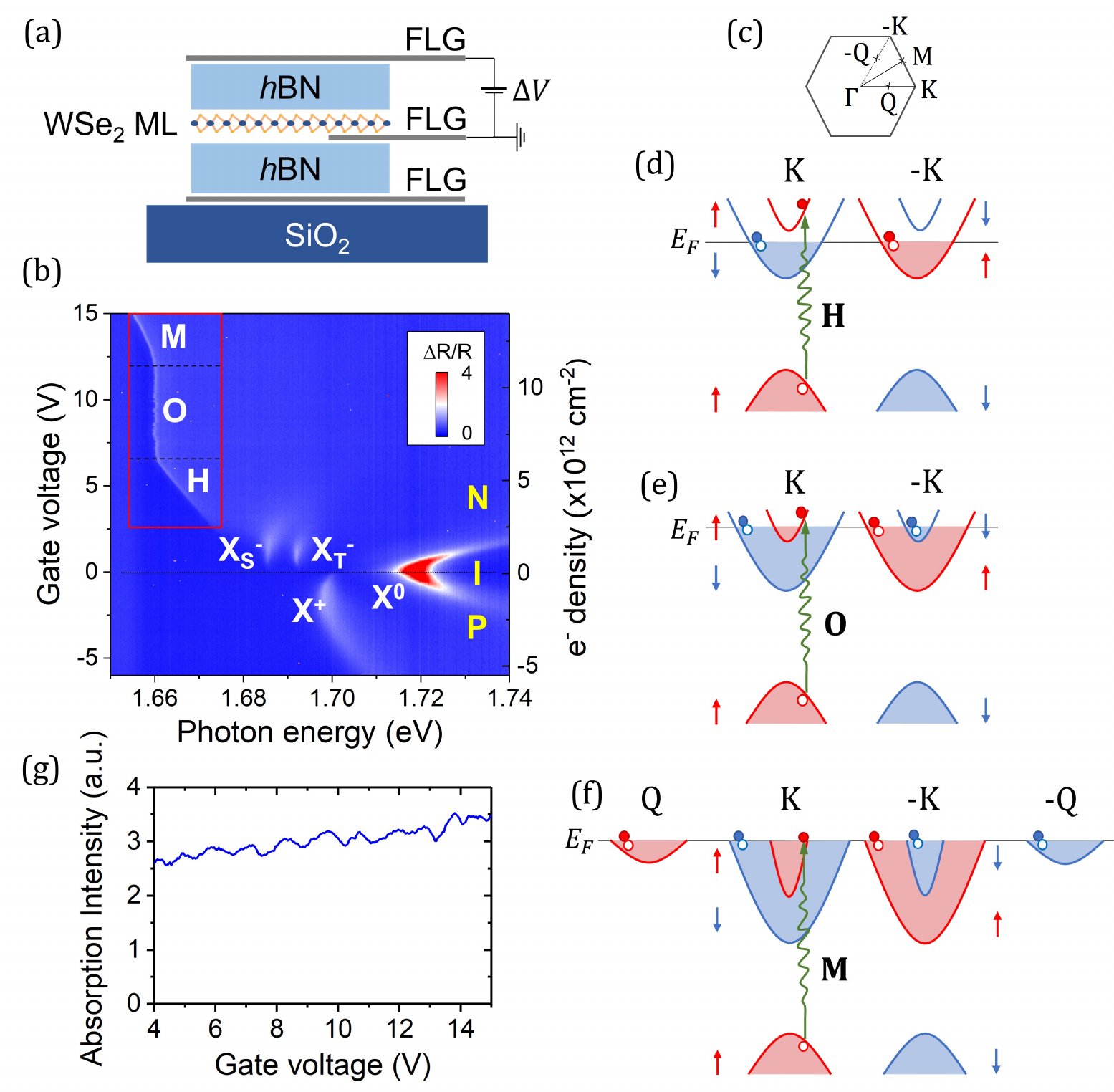}
\caption{(a) Sketch of the sample. (b) Reflectivity contrast as a function of the gate voltage. (c) 1st Brillouin zone of WSe$_2$. Sketches of (d) hexciton $H$, (e) oxciton $O$, and (f) multi-valley complex $M$. (g) Spectrally integrated intensity of the reflectivity contrast as a function of the gate voltage.}
\label{fig1}
\end{figure}

The sample structure investigated in this study is illustrated in Fig.\ref{fig1}a. A WSe$_2$ monolayer is embedded in a dual-gate, charge-tunable device. Details of the fabrication process are provided in the Supplemental Material. In the experiments reported here, only the top gate is used: a bias voltage is applied between a few-layer graphene (FLG) electrode and the WSe$_2$ monolayer, separated by a 19 nm-thick hBN dielectric. The monolayer is grounded so that a positive gate voltage induces electrostatic $n$-doping in WSe$_2$. The breakdown field of hBN has been reported to be around 0.5 V.nm$^{-1}$ \citeA{pierret_dielectric_2022}. But remarkably, we are able to increase the gate voltage up to 15 V (corresponding to an electric field of 0.8 V.nm$^{-1}$ across the hBN layer) without detecting any leakage current above 1 nA. This corresponds to an electron density of $1.4\times 10^{13}$ cm$^{-2}$. At higher biases, we observe a photocurrent induced by laser excitation probably in the graphite electrodes, and therefore restrict our measurements to voltages below 15 V.

Optical experiments are carried out at 4 K in a closed-cycle helium cryostat. Continuous-wave (cw) PL is performed using the 633 nm (1.96~eV) line of a HeNe laser focused through a high-numerical-aperture objective (NA = 0.82). Reflectivity spectra are recorded under tungsten–halogen white-light illumination. The PL is spectrally dispersed in a monochromator and detected with a charge coupled device (CCD) camera. TRPL measurements are performed by exciting the sample at 635 nm with picosecond pulses from a Ti:Sapphire laser pumping an optical parametric oscillator (OPO). The PL dynamics are analysed with a streak camera coupled to a monochromator; for improved temporal resolution (1~ps for the half width at half maximum of the instrument response function shown in Fig.~\ref{fig3}a), the grating is replaced by a mirror, while spectral filtering is ensured using tunable long-pass and short-pass filters. 

In all experiments, the excitation spot diameter is $\sim$1~µm, the excitation wavelength lies above the free-carrier band gap at 1.89~eV \citeA{kapuscinski_rydberg_2021, stier_magnetooptics_2018}, and the excitation power is kept at low levels (1 µW for cw PL and 100 nW for TRPL). We checked that lower excitation powers do not change the results of this work (see Supplemental Material).

Figure~\ref{fig1}b presents the reflectivity spectra, defined as  
\[
\Delta R/R=\frac{R_{\mathrm{flake}}-R_{\mathrm{ref}}}{R_{\mathrm{ref}}},
\]  
as a function of electron density, where $R_{\mathrm{flake}}$ is the reflectivity measured on the WSe$_2$ monolayer and $R_{\mathrm{ref}}$ is the reflectivity of the underlying structure without WSe$_2$ (the extrapolation procedure for $R_{\mathrm{ref}}$ is described in the Supplemental Material). We first identify the well-documented neutral exciton resonance $X^{0}$ at 1.72~eV, which blueshifts and gradually disappears as the doping density increases. At low doping levels (a few $10^{11}$~cm$^{-2}$), the charged exciton resonances $X^{+}$ (positive bright trion), $X_{S}^{-}$ (negative bright singlet trion), and $X_{T}^{-}$ (negative bright triplet trion) are observed, consistent with previous reports \citeA{courtade_charged_2017}.  

The focus of this study is the high electron doping regime, where the $X_{S}^{-}$ and $X_{T}^{-}$ resonances vanish and new, lower-energy transitions emerge below 1.67~eV. In this regime, the simple three-particle trion picture is no longer valid, and the interaction between excitons and the Fermi seas in different valleys must be taken into account. Several theoretical frameworks have been proposed to describe this interaction, including exciton--Fermi polarons \citeA{efimkin_many-body_2017, sidler_fermi_2017, glazov_optical_2020} and correlated trions \citeA{suris_excitons_2001, bronold_absorption_2000, chang_crossover_2018}. The low energy transitions labeled $H$ and $O$ in Fig.~\ref{fig1}b have been observed in previous works \citeA{jones_optical_2013, wang_probing_2017} but their identification remained elusive until recently. Li \textit{et al.} first demonstrated that these transitions involve many-body exciton states with intervalley correlations \citeA{li_many-body_2022}. More recently, magneto-reflectivity studies have identified these transitions as six and eight particle complexes \citeA{van_tuan_six-body_2022}. Schematic illustrations of these complexes are shown in Fig.~\ref{fig1}d,e.  

In WSe$_2$ monolayers, the direct band gap is located at the $K$ points of the Brillouin zone, and the conduction band is split by spin--orbit coupling with $\Delta_{SO}^{CB}\sim 12$~meV \citeA{kapuscinski_rydberg_2021, ren_measurement_2023, jindal_brightened_2025}. A bright optical transition requires spin conservation within the photogenerated electron--hole pair, which in WSe$_2$ occurs when an electron from the top valence band is promoted to the second conduction band. At low doping, the Fermi level populates the first conduction band. At sufficiently high electron density, however, the photogenerated exciton (for instance at the $K$ valley) can bind to one electron from the Fermi sea in the bottom $K$ valley and another in the bottom $-K$ valley, leaving behind two Fermi holes. This six-particle complex has been termed the \textit{hexciton} ($H$). When the Fermi level rises into the second conduction band, the complex can further bind to an additional electron from the top $-K$ valley and its corresponding Fermi hole, forming an eight-particle state known as the \textit{oxciton} ($O$).  
The transition from $H$ to $O$ is clearly observed in Fig.~\ref{fig1}b as a change in the slope of the energy shift: while $H$ redshifts with increasing doping, the $O$ resonance remains nearly constant. This behavior has been recently connected to the concept of particles distinguishability, as discussed in \citeA{dery_energy_2025}. In particular, the energy evolution arises from a competition between three mechanisms: screening-induced reduction of the binding energy, band-gap renormalization, and shake-up processes.  

At even higher doping densities ($n \approx 1.2\times 10^{13}$~cm$^{-2}$), we observe the emergence of an additional redshifting resonance, labeled $M$ in Fig.~\ref{fig1}b. This feature, only very recently reported by Dijkstra \textit{et al.} \citeA{dijkstra_ten-valley_2025}, has been attributed to a multi-valley excitonic complex involving carriers from the $K$, $-K$, and the six $Q$ valleys situated between the $K$ and $\Gamma$ points (Fig.~\ref{fig1}c), predicted to lie $\Delta_{KQ}\sim 35$~meV above the $K$ valley in WSe$_2$ monolayers (Fig.~\ref{fig1}f) \citeA{kormanyos_kp_2015}.

\begin{figure}[h!]
\includegraphics[width=\linewidth]{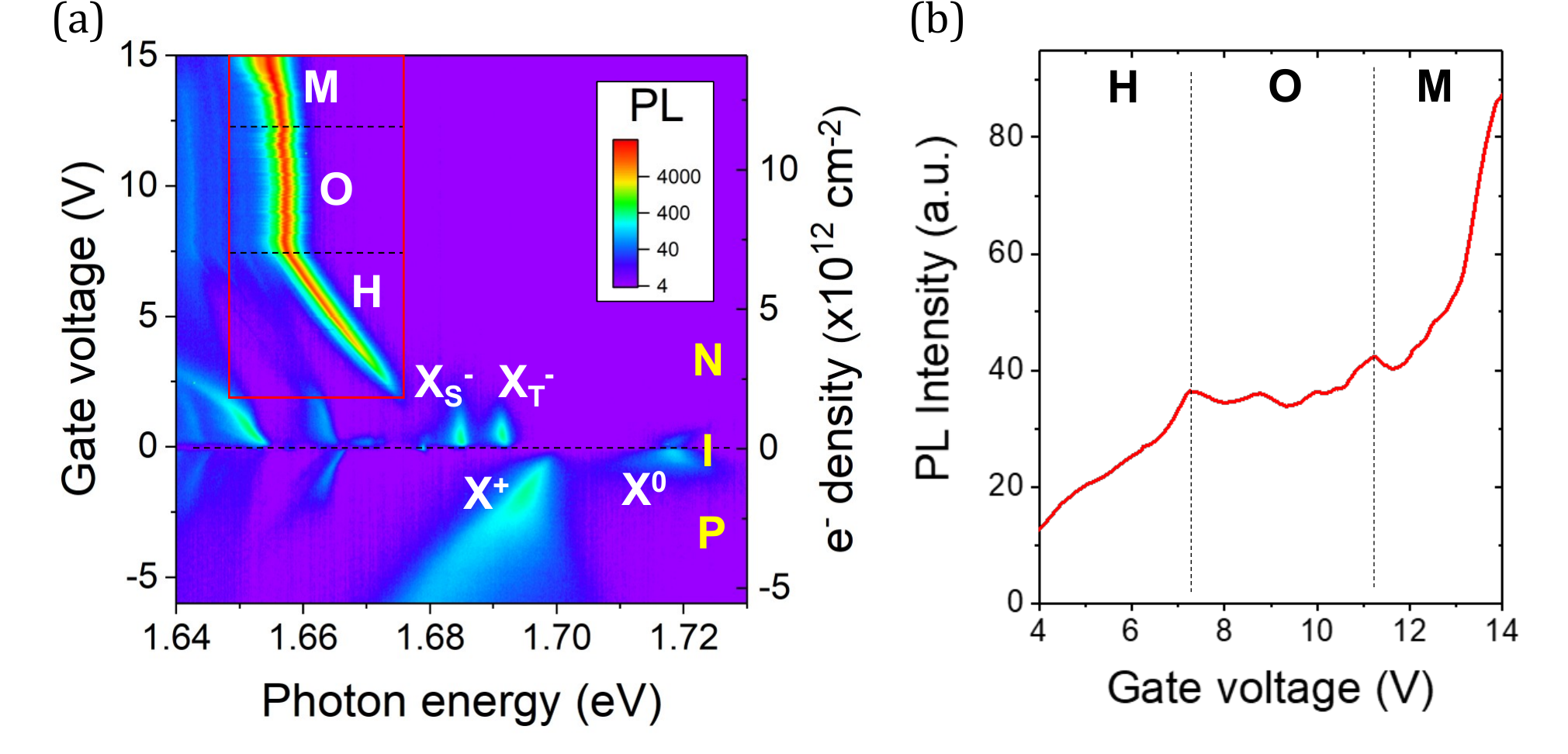}
\caption{(a) cw PL intensity as a function of the gate voltage. (b) Spectrally integrated PL intensity of $H$, $O$ and $M$ peaks as a function of gate voltage.}
\label{fig2}
\end{figure}

\begin{figure*}[t!]
\includegraphics[width=\linewidth]{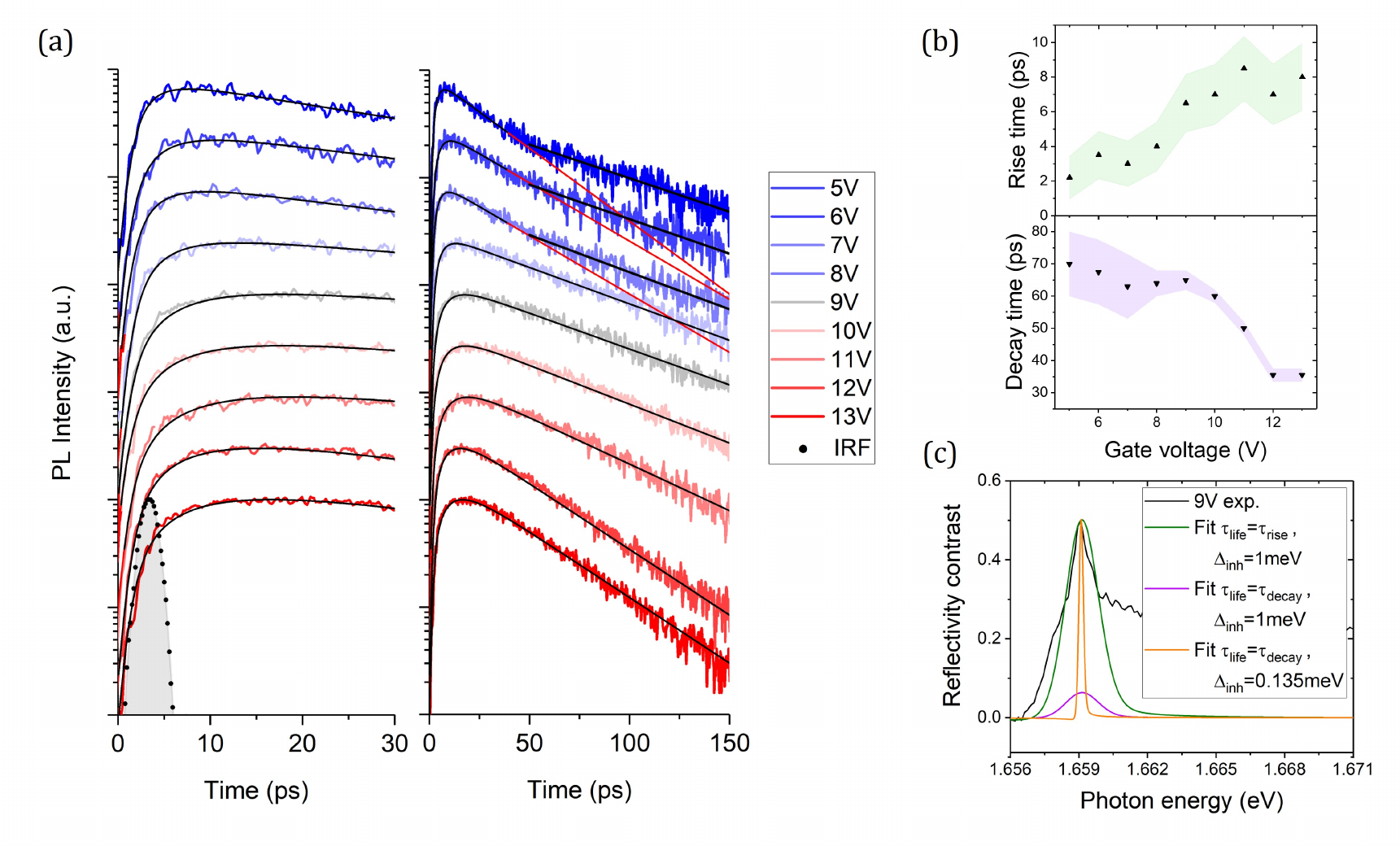}
\caption{(a) Normalized TRPL as a function of the gate voltage. The curves are shifted vertically for clarity. The rise times (decay times) are extracted from the data on the left (right) panel with two different time resolutions. Fits are shown by the thin black solid lines. Between 5V and 7V, the decay time is not monoexponential. The red lines represent the short decay time contribution. See the Supplemental Material for more details. (b) Rise time $\tau_\mathrm{rise}$ and decay time $\tau_\mathrm{decay}$ as a function of gate voltage. Coloured shaded area represent the error bars. (c) Reflectivity contrast at 9 V (black solid line) and result of the transfer matrix model for $\tau_\mathrm{life}\equiv\tau_\mathrm{rise}$ (green line) and $\tau_\mathrm{life}\equiv\tau_\mathrm{decay}$ (yellow and purple lines).
}
\label{fig3}
\end{figure*}

We now present in Figure~\ref{fig2}a the cw PL spectra as a function of the doping density. Near charge neutrality and in the low-doping regime, we clearly identify the bright transitions ($X^{0}$, $X^{+}$, $X_{S}^{-}$ and $X_{T}^{-}$), as well as several low-energy features associated with dark excitons and their phonon replicas, which are absent in reflectivity spectra due to their weak oscillator strength. Remarkably, in the highly $n$-doped regime, the multi-particle excitonic complexes $H$, $O$, and $M$ give rise to a strong PL signal, up to two orders of magnitude more intense than any other transition.
Fig.~\ref{fig2}b displays the integrated PL intensity of these complexes as a function of the gate voltage. The signal increases significantly in the $H$ and $M$ regimes, while remaining nearly constant in the $O$ regime. This behavior contrasts with the reflectivity data, which show only a weak increase in the integrated intensity of the reflectivity contrast with doping (Fig.~\ref{fig1}g), suggesting that the oscillator strength itself undergoes only minor changes.  

\begin{figure}[h!]
\includegraphics[width=\linewidth]{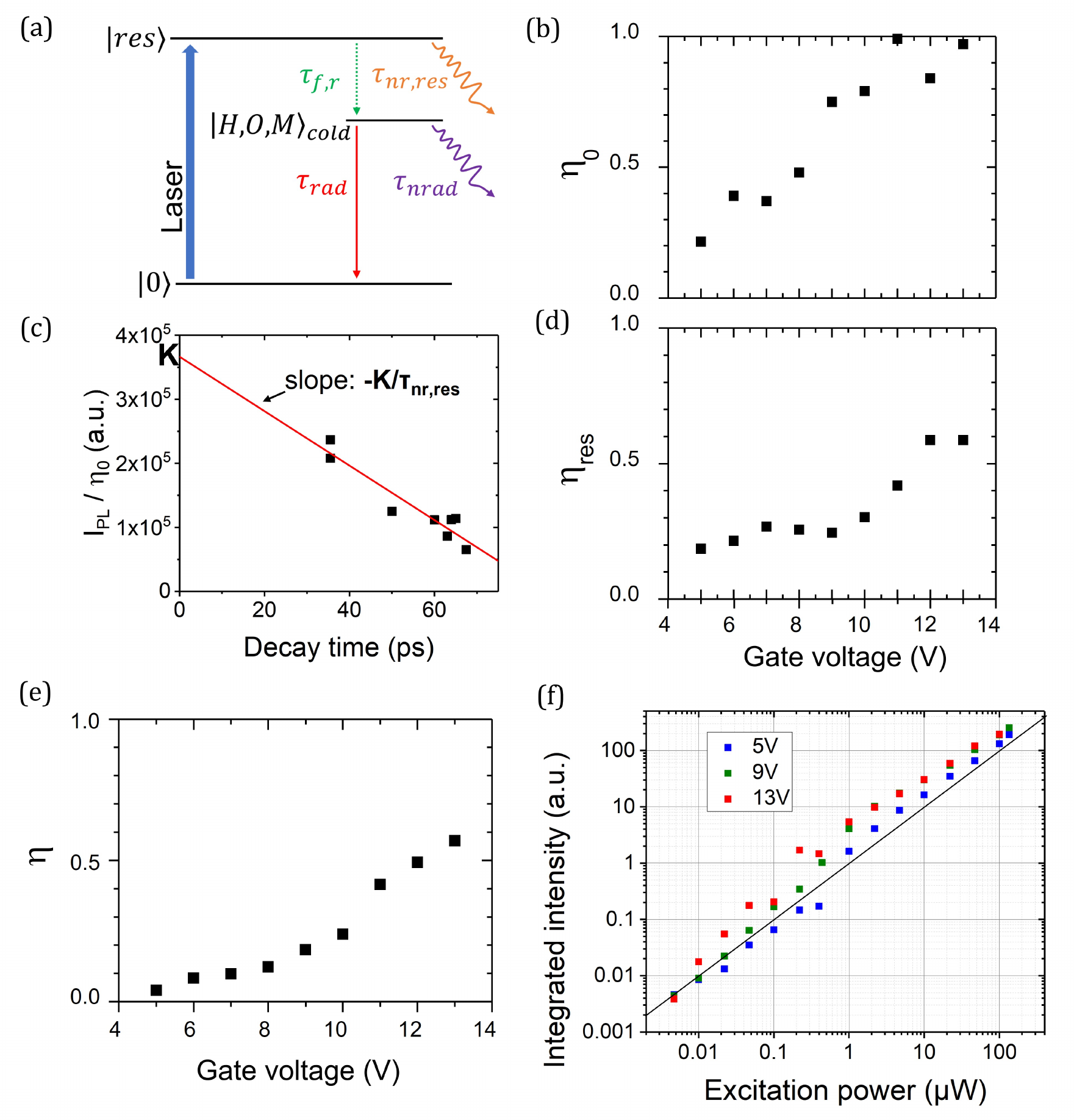}
\caption{(a) Sketch of the three-level model.  (e) Total quantum yield $\eta$ and its components $\eta_{0}$ (b) and $\eta_\mathrm{res}$ (d) as a function of the gate voltage. (c) $I_{PL}/\eta_{0}$ as a function of $\tau_\mathrm{decay}$. (f) PL intensity as a function of the gate voltage for the three regimes $H$ (5V), $O$ (9V) and $M$ (13V) in cw excitation.}
\label{fig4}
\end{figure}

To gain further insight into this behavior, we performed TRPL measurements in the $H$, $O$, and $M$ regimes (Fig.~\ref{fig3}a). Except at the lowest voltages (5--7 V), the dynamics are well described by a bi-exponential response, characterized by a rise time $\tau_\mathrm{rise}$ of a few ps (resolvable thanks to our time resolution) and a decay time $\tau_\mathrm{decay}$ of a few tens of ps:  
\begin{equation}
I_\mathrm{PL}(t)\propto \exp(-t/\tau_\mathrm{decay})-\exp(-t/\tau_\mathrm{rise}).
\label{IPL_decay_rise}
\end{equation}
Details of the fitting procedure are provided in the Supplemental Material. The extracted rise and decay times are plotted in Fig.~\ref{fig3}b as a function of the gate voltage, revealing that $\tau_\mathrm{rise}$ increases whereas $\tau_\mathrm{decay}$ decreases when the doping density increases.  
To rationalize this dynamics, we introduce the simple three-level model sketched in Fig.~\ref{fig4}a. Under non-resonant excitation (above-band-gap pumping), electron--hole pairs are first photogenerated in a reservoir state. A fraction of these carriers recombines non-radiatively with a characteristic time $\tau_\mathrm{nr,res}$, while the remainder form and relax into the bound many-body complexes $H$, $O$, or $M$ with a formation/relaxation time $\tau_\mathrm{f,r}$. These cold complexes, denoted $\ket{H,O,M}_\mathrm{cold}$, can then recombine either radiatively with lifetime $\tau_\mathrm{rad}$ or non-radiatively with lifetime $\tau_\mathrm{nrad}$. The effective lifetime of the complexes and the reservoir are thus defined as:  
\begin{equation}
\tau_\mathrm{life}^{-1}=\tau_\mathrm{rad}^{-1}+\tau_\mathrm{nrad}^{-1},
\label{taulife}
\end{equation}
\begin{equation}
\tau_\mathrm{res}^{-1}=\tau_\mathrm{f,r}^{-1}+\tau_\mathrm{nr,res}^{-1}.
\label{taures}
\end{equation}
Although minimal, this three-level model includes all relevant relaxation pathways. The microscopic details of the formation and relaxation processes entering $\tau_\mathrm{f,r}$ are beyond the scope of this work.  

Within this framework, the PL dynamics can be expressed as \citeA{fang_control_2019}:  
\begin{equation}
    I_\mathrm{PL}(t)\propto \frac{1}{\tau_\mathrm{res}-\tau_\mathrm{life}} \left[ \exp\left(-\frac{t}{\tau_\mathrm{res}}\right)-\exp\left(-\frac{t}{\tau_\mathrm{life}}\right) \right]
    \label{IPL}
\end{equation}
We see from Eq.~\ref{IPL_decay_rise} and Eq.~\ref{IPL} that the correspondence between $\tau_\mathrm{rise}$ and $\tau_\mathrm{decay}$ measured in TRPL and $\tau_\mathrm{res}$ and $\tau_\mathrm{life}$ from the model depends on the sign of $\tau_\mathrm{res}-\tau_\mathrm{life}$. If $\tau_\mathrm{res}<\tau_\mathrm{life}$, then $\tau_\mathrm{res}\equiv\tau_\mathrm{rise}$ and $\tau_\mathrm{life}\equiv\tau_\mathrm{decay}$. On the opposite, if $\tau_\mathrm{life}<\tau_\mathrm{res}$, then $\tau_\mathrm{life}\equiv\tau_\mathrm{rise}$ and $\tau_\mathrm{res}\equiv\tau_\mathrm{decay}$.
TRPL alone cannot distinguish between these two scenarios. A full understanding requires analyzing the reflectivity contrast—both its shape and amplitude—which depend intimately on the radiative and non-radiative lifetimes.

Fig.~\ref{fig3}c shows the reflectivity contrast measured at 9 V (i.e., in the $O$ regime). The spectrum consists of a peak associated with the $O$ transition and a high-energy tail of unknown origin. The peak is well described using a transfer-matrix model \citeA{robert_optical_2018}, which accounts for multiple reflections in the heterostructure and treats the excitonic resonance as a Lorentz oscillator convoluted with a Gaussian. The Lorentzian linewidth contains both a radiative component ($\propto \tau_\mathrm{rad}^{-1}$) and a non-radiative component ($\propto \tau_\mathrm{nrad}^{-1}$), while the Gaussian broadening $\Delta_\mathrm{inh}$ accounts for the inhomogeneous disorder. Importantly, the number of free parameters is reduced to two, since $\tau_\mathrm{rad}$ and $\tau_\mathrm{nrad}$ are constrained by Eq.~\ref{taulife}, where $\tau_\mathrm{life}$ is directly obtained from TRPL ($\tau_\mathrm{life}\equiv\tau_\mathrm{decay}$ or $\tau_\mathrm{life}\equiv\tau_\mathrm{rise}$). The two independent fitting parameters are thus $\Delta_\mathrm{inh}$ and $\eta_{0}=\tau_\mathrm{life}/\tau_\mathrm{rad}$.  

The results of the model are presented in Fig.~\ref{fig3}c for both cases. When $\tau_\mathrm{life}\equiv\tau_\mathrm{rise}$, excellent agreement is obtained for both amplitude and linewidth using $\eta_{0}=0.75$ and $\Delta_\mathrm{inh}=1$~meV, consistent with the PL linewidth. In contrast, assuming $\tau_\mathrm{life}\equiv\tau_\mathrm{decay}$ fails to reproduce the experimental data for any reasonable set of parameters. For example, calculations with $\eta_{0}=1$ and $\Delta_\mathrm{inh}=1$~meV (purple line) or with $\eta_{0}=1$ and $\Delta_\mathrm{inh}=0.135$~meV (orange line) cannot simultaneously fit both amplitude and linewidth.  

We therefore conclude that $\tau_\mathrm{life}\equiv\tau_\mathrm{rise}$ and $\tau_\mathrm{res}\equiv\tau_\mathrm{decay}$. This represents an unusual situation where $\tau_\mathrm{life}<\tau_\mathrm{res}$ (\textit{i.e.} where the recombination time is faster than the formation and relaxation time of the complex), but similar behavior has already been reported for the neutral exciton dynamics in MoSe$_{2}$ \citeA{fang_control_2019}.  

We now turn to the extraction of the quantum yield $\eta$ from our data. Within the framework of the model in Fig.~\ref{fig4}a, it can be expressed as  
\begin{equation}
\eta = \eta_{0} \times \eta_\mathrm{res} 
     = \frac{\tau_\mathrm{life}}{\tau_\mathrm{rad}} \times \frac{\tau_\mathrm{res}}{\tau_\mathrm{f,r}} ,
\label{QY_general}
\end{equation}
where $\eta_{0}$ accounts for the balance between radiative and non-radiative channels of cold complexes, and $\eta_\mathrm{res}$ quantifies the efficiency of relaxation from the reservoir into the cold complexes.  

The parameter $\eta_{0}$ can be extracted from the combined analysis of reflectivity and TRPL, as described above. Its evolution with gate voltage is shown in Fig.~\ref{fig4}b (with detailed fits given in the Supplemental Material). We find that $\eta_{0}$ approaches unity in the $M$ regime, indicating that $\tau_\mathrm{rad}\ll \tau_\mathrm{nrad}$. We interpret this result as a consequence of the quenching of non-radiative and dark channels. When the Fermi level crosses the second conduction band, the ground state becomes the oxciton (\textit{i.e.} a bright state), whereas in the neutral and moderately $n$-doped regimes it corresponds to a dark exciton or dark trion.

To determine $\eta_\mathrm{res}$, we exploit the fact that the PL intensity is proportional to the total quantum yield $\eta$. It can thus be written as:  
\begin{equation}
I_\mathrm{PL} = K \times \eta 
              = K \times \eta_{0} \times \eta_\mathrm{res} 
              = K \times \eta_{0} \times \left( 1 - \frac{\tau_\mathrm{decay}}{\tau_\mathrm{nr,res}} \right),
\label{QY_PL}
\end{equation}
where $K$ is an experimental proportionality factor that depends on the absorption coefficient and the optical setup. We assume $K$ to be independent of the gate voltage (\textit{i.e.} the absorption is constant with the gate voltage), as supported by reflectivity measurements at the excitation wavelength, which show no gate-induced variation (see Supplemental Material).  

In Fig.~\ref{fig4}c we plot $I_\mathrm{PL}/\eta_{0}$ as a function of $\tau_\mathrm{decay}$. The data fall on a straight line, as expected from Eq.~\ref{QY_PL}, from which we extract both $K$ (intercept) and $\tau_\mathrm{nr,res}$ (from the slope). We obtain $\tau_\mathrm{nr,res}$ = 86 ps. The linear dependence indicates that $\tau_\mathrm{nr,res}$ is essentially independent of the gate voltage.  

Using this value together with $\tau_\mathrm{decay} \equiv \tau_\mathrm{res}$, we  extract $\tau_\mathrm{f,r}$ from Eq.~\ref{taures}, and subsequently determine $\eta_\mathrm{res}$ as a function of gate voltage. The results, presented in Fig.~\ref{fig4}d, show that $\eta_\mathrm{res}$ increases with doping, indicating that the formation of cold complexes becomes more efficient at higher carrier densities. As shown in Fig.~\ref{fig4}e, the total quantum yield $\eta$ also rises steadily with doping, reaching values as high as 60\% at $V_\mathrm{g}=13$ V.  

Finally, we present in Fig.~\ref{fig4}f the variation of the spectrally integrated cw-PL intensity with the excitation power in the three regimes $H$, $O$, and $M$. Remarkably, the PL intensity is perfectly linear with power in the three regimes.

In conclusion, we have investigated many-body excitonic complexes emerging in heavily $n$-doped WSe$_2$ monolayers. In this high-density regime, we observe a pronounced enhancement of the PL signal, which increases with doping across both the $H$ and $M$ regimes. By combining reflectivity and TRPL measurements, we demonstrate that the quantum yield reaches remarkably high values, in sharp contrast to the characteristically low efficiencies of the neutral and low-density regimes. These results underscore the potential of TMD monolayers as a platform for exploring excitons interacting with dense electron gases—a regime inaccessible in conventional quantum wells due to their lower Mott transition threshold. Moreover, the high quantum yield and the linear dependence of the emission intensity on excitation power open promising avenues for the design of efficient light-emitting devices based on TMD monolayers.
\newline

\textit{Aknowledgements:}
We thank S.A. Crooker for fruitful discussions and M.M. Glazov for helping us developing the transfer matrix formalism. This work was supported by Agence Nationale de la Recherche funding under the ANR ATOEMS and ANR IXTASE, the grant NanoX n° ANR-17-EURE-0009 in the framework of the "Programme des Investissements d’Avenir". K.W. and T.T. acknowledge support from the JSPS KAKENHI (Grant Numbers 21H05233 and 23H02052), the CREST (JPMJCR24A5), JST and World Premier International Research Center Initiative (WPI), MEXT, Japan. Work at the University of Rochester was supported by the Department of Energy, Basic Energy Sciences, Division of Materials Sciences and Engineering under Award No. DE-SC0014349. The sample has been fabricated using the Exfolab platform.

\bibliographystyleA{unsrt}
\bibliographyA{}

\newpage
\onecolumngrid

\setcounter{figure}{0}
\renewcommand{\figurename}{Fig.}
\renewcommand{\thefigure}{S\arabic{figure}}

\begin{center}
{\large \bfseries Supplemental Material: \\[0.5em]
High luminescence efficiency of multi-valley excitonic complexes in heavily doped WSe$_2$ monolayer}

\vspace{1em}

Sébastien Roux,$^{1,*}$ Tilly Guyot,$^{1,*}$ Abraaao Cefas Torres-Dias,$^{1}$ Delphine \\ Lagarde,$^{1}$ Laurent Lombez,$^{1}$ Dinh Van Tuan,$^{2}$ Junghwan Kim,$^{2}$ Kenji Watanabe,$^{3}$\\ Xavier Marie,$^{1,4}$ Takashi Taniguchi,$^{5}$ Hanan Dery,$^{2,6}$ and Cedric Robert$^{1,\dagger}$

\vspace{0.5em}

\small
\textit{$^1$Université de Toulouse, INSA-CNRS-UPS, LPCNO, 135 Av. Rangueil, 31077 Toulouse, France\\}
\textit{$^2$Department of Electrical and Computer Engineering, \\University of Rochester, Rochester, NY, United States\\}
\textit{$^3$Research Center for Electronic and Optical Materials, \\National Institute for Materials Science, 1-1 Namiki, Tsukuba 305-0044, Japan\\}
\textit{$^4$Institut Universitaire de France, 75231, Paris, France\\}
\textit{$^5$Research Center for Materials Nanoarchitectonics,\\ National Institute for Materials Science, 1-1 Namiki, Tsukuba 305-0044, Japan\\}
\textit{$^6$Department of Physics and Astronomy, University of Rochester, Rochester, NY, United States}

\vspace{2em}
\end{center}

\subsection{Sample fabrication}
The hBN flakes were obtained by exfoliating a bulk crystal synthesized using the reference method under high pressure and high temperature \citeB{taniguchi2007synthesis}. The few layer graphene flakes were obtained from a Highly Ordered Pyrolitic Graphite (HOPG) crystal from HQ Graphene and the WSe$_2$ monolayer was exfoliated from a bulk crystal from 2D Semiconductor. The flakes were exfoliated with a PDMS stamp and transferred to a Si/SiO$_2$ (83nm thick) substrate by the dry transfer method described in Ref. \citeB{castellanos2014deterministic}. Then, photolithography and metal deposition (Cr/Au) was performed to connect the graphite gates and the WSe$_2$ monolayer to large pads. The thickness of the flakes were measured by atomic force microscopy and optical contrast.

\subsection{Power dependent experiments}

\begin{figure}[h!]
\includegraphics[scale=0.53]{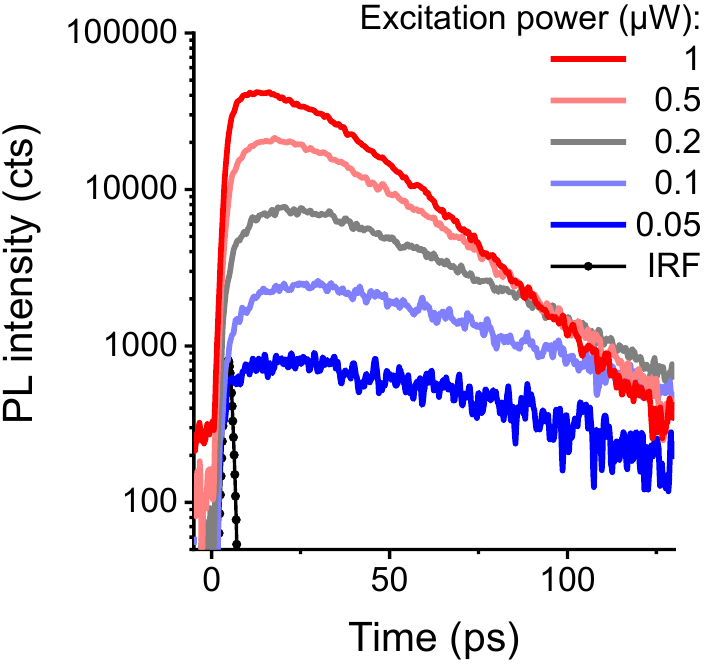}
\caption{TRPL measurements for different excitation powers measured at 9 V after pulse excitation at 635 nm at $t = 0$.}
\label{S1}
\end{figure}

Figure S1 shows the time-resolved photoluminescence (TRPL) intensity at 9 V for various excitation powers. At low power, the decay profiles are identical, but they accelerate as the power increases. To mitigate any power-dependent effects, all TRPL measurements in this study were conducted at a constant excitation power of 100 nW. The changes in decay dynamics at higher powers are beyond the scope of this paper and warrant further investigation in future work.

\subsection{Differential reflectivity}
Figure S2 shows the raw reflectivity spectra measured between -8 V and 12 V. At the excitation energy of 1.96 eV used in PL and TRPL experiments, the reflectivity is independent of voltage, confirming that the absorption at this energy is doping-independent.\\\par

To calculate the reflectivity contrast (shown in Figure 1b of the main text), we need to determine \( R_{\mathrm{ref}} \), the reflectivity of the full structure without the WSe\(_2\) monolayer (ML). Usually, this is done by measuring the reflectivity in a region of the sample free from WSe\(_2\). However, this method can introduce uncertainties because slight movements across the sample can shift the focal point. To minimize this effect, we instead calculate \( R_{\mathrm{ref}} \) using the gate-dependent spectra taken from the WSe\(_2\) ML itself. At both low and high energies, the spectra are identical, and no excitonic transitions occur, allowing us to average these spectra to obtain \( R_{\mathrm{ref}} \). For the energy range where excitonic transitions are present, we interpolate \( R_{\mathrm{ref}} \) using a polynomial function. The resulting \( R_{\mathrm{ref}} \) spectrum is shown as the black line in Figure S2.

\begin{figure}[h!]
\includegraphics[scale=0.25]{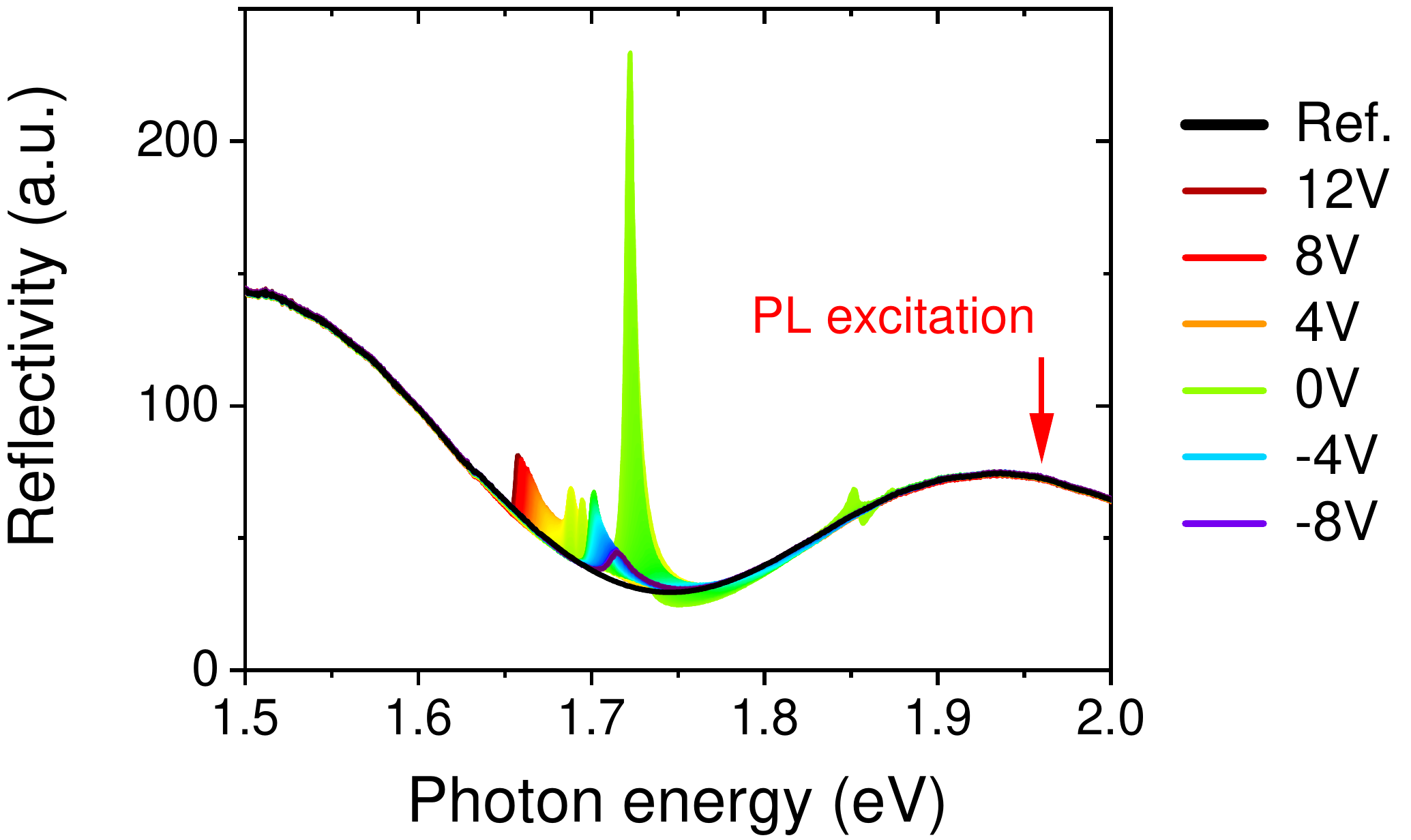}
\caption{Raw reflectivity spectra.}
\label{S2}
\end{figure}

\subsection{Transfer matrix model for calculating reflectivity contrast}

\noindent To simulate the reflectivity contrast, we rely on the transfer matrix formalism described in \citeB{robert_optical_2018}.\\\par

\noindent The transfer matrix for a layer of thickness $L$ is defined as:
\begin{equation}
    \hat{T} =
    \begin{pmatrix}
    \exp  \left(\mathrm i kL \right) & 0 \\
    0 & \exp  \left(-\mathrm i kL\right) 
    \end{pmatrix}
\end{equation}
with $k=\omega n/c$, n is the refractive index of the layer.\\\par

\noindent For an interface between two layers of index $n_1$ and $n_2$, the transfer matrix is written as:
\begin{equation}
    \hat{T}_{n_1 \rightarrow n_2} = \frac{1}{2n_1}
    \begin{pmatrix}
    n_1+n_2 & n_2-n_1 \\
    n_2-n_1 & n_1+n_2 
    \end{pmatrix}
\end{equation}
The refractive indexes used for each material are summarized in Table~\ref{table_index}

\begin{table}[h!]
\centering
\begin{tabular}{|c|c|c|c|c|c|}
\hline
\rowcolor[HTML]{EFEFEF} 
Material                                 & vacuum & hBN & SiO$_2$ & Si & FLG \\ \hline
\cellcolor[HTML]{EFEFEF}Refractive index & 1         & 2.2       & 1.46       & 3.5 & 2.57+1.3i      \\ \hline
\end{tabular}
\caption{Refractive indexes used for the calculations.}
\label{table_index}
\end{table}

\noindent For the WSe$_2$ ML, the transfer matrix is:
\begin{equation}
    \hat{T}_{ML} = \frac{1}{t}
    \begin{pmatrix}
    t^2-r^2 & r \\
    -r & 1 
    \end{pmatrix}
\end{equation}
where $r$ and $t$ are the reflection and transmission coefficients of the ML.
We have:
\begin{equation}
    r(E)=\int_0^\infty dx\frac{1}{\sqrt{\pi\Delta_{\mathrm{inh}}}} \exp{\left[ - \left( \frac{x-E_0}{\Delta_{\mathrm{inh}}} \right)^2\right]} \times \frac{-i\Gamma_{\mathrm{rad}}^{\mathrm{vac}}}{E-x+i\left( \Gamma_{\mathrm{rad}}^{\mathrm{vac}} +\Gamma_{\mathrm{nrad}}\right)} \quad \mathrm{and} \quad t=1+r
\label{r}
\end{equation}
where $E$ is the energy of the reflected photon and $E_0$ the energy of the transition. $\Delta_{\mathrm{inh}}$ is the inhomogeneous broadening (the full width at half maximum (FWHM) is $2\sqrt{\ln{2}}\Delta_{\mathrm{inh}}$).
$\Gamma_{\mathrm{nrad}}$ is the non-radiative linewidth (half width at half maximum (HWHM)) and $\Gamma_{\mathrm{rad}}^{\mathrm{vac}}$ is the radiative linewidth in vacuum (HWHM).\\\par

\noindent We define the homogeneous linewidth (HWHM) in the heterostructure as:
\begin{equation}
    \Gamma=\frac{\hbar}{2\tau_{\mathrm{life}}}
\end{equation}
The non-radiative linewidth is defined by the quantum yield $\eta_0$ and $\Gamma$:
\begin{equation}
    \Gamma_{\mathrm{nrad}}=(1-\eta_0) \times \Gamma
\end{equation}
The radiative linewidth in the heterostructure is:
\begin{equation}
    \Gamma_{\mathrm{rad}}^{\mathrm{het}}=\eta_0 \times \Gamma
\end{equation}
with
\begin{equation}
    \Gamma_{\mathrm{rad}}^{\mathrm{het}}=\frac{\hbar}{2\tau_{\mathrm{rad}}}
\end{equation}
$\Gamma_{\mathrm{rad}}^{\mathrm{het}}$ and the radiative linewidth in infinite homogeneous hBN $\Gamma_{\mathrm{rad}}^{\mathrm{hBN}}$ are linked by the Purcell factor $\mathcal{F}_P$:
\begin{equation}
    \Gamma_{\mathrm{rad}}^{\mathrm{het}}=\Gamma_{\mathrm{rad}}^{\mathrm{hBN}} \times \mathcal{F}_P
\end{equation}
The Purcell factor is calculated using the transfer matrix formalism. We find $\mathcal{F}_P=1.46$ for our structure.\\\par

\noindent Finally, $\Gamma_{\mathrm{rad}}^{\mathrm{hBN}}$ and $\Gamma_{\mathrm{rad}}^{\mathrm{vac}}$ in Eq.\ref{r} are linked by:
\begin{equation}
    \Gamma_{\mathrm{rad}}^{\mathrm{vac}}=\Gamma_{\mathrm{rad}}^{\mathrm{hBN}} \times n_{\mathrm{hBN}}
\end{equation}
where $n_{\mathrm{hBN}}$ is the refractive index of hBN.\\\par

\noindent The transfer matrix of the full heterostructure is:

\begin{align}
\hat{T}_\mathrm{tot}
  &=\hat{T}_{\mathrm{SiO_2 \rightarrow Si}}\hat{T}_{\mathrm{SiO_2}}\hat{T}_{\mathrm{FLG \rightarrow SiO_2}}\hat{T}_{\mathrm{bottom~FLG}}\hat{T}_{\mathrm{hBN \rightarrow FLG}}\hat{T}_{\mathrm{bottom~hBN}} \notag \\[6pt]
  &\quad \times \hat{T}_{\mathrm{vac \rightarrow hBN}}\hat{T}_{ML}\hat{T}_{\mathrm{hBN \rightarrow vac}}\hat{T}_{\mathrm{top~hBN}}\hat{T}_{\mathrm{FLG \rightarrow hBN}}\hat{T}_{\mathrm{top~FLG}}\hat{T}_{\mathrm{vac \rightarrow FLG}}
\end{align}~\\\par

\noindent And the reflectivity on the full structure is:
\begin{equation}
    R_{\mathrm{flake}}=\left | -\frac{\hat{T}_{\mathrm{tot}}\left(2,1 \right)}{\hat{T}_{\mathrm{tot}}\left(2,2 \right)} \right |^2
\end{equation}
The reference reflectivity $R_{\mathrm{ref}}$ is calculated in the same way by replacing $\hat{T}_{ML}$ by a unity matrix.

\subsection{Fitting of the reflectivity constrast}

The differential reflectivity spectra are shown in Fig. S3. The $H$, $O$, and $M$ transitions are fitted using the transfer-matrix model described in the previous section from which the value of the parameter $\eta_{0}$ is extracted. In practice, we fix the parameter $\Delta_\mathrm{inh}$=1~meV in agreement with the measured linewidth in PL (the linewidth is dominated by inhomogeneous broadening) and we use the TRPL measurement of $\tau_{\mathrm{life}}$ in our model so that the only fitting parameter is $\eta_{0}$.

\begin{figure}[h!]
\includegraphics[width=\linewidth]{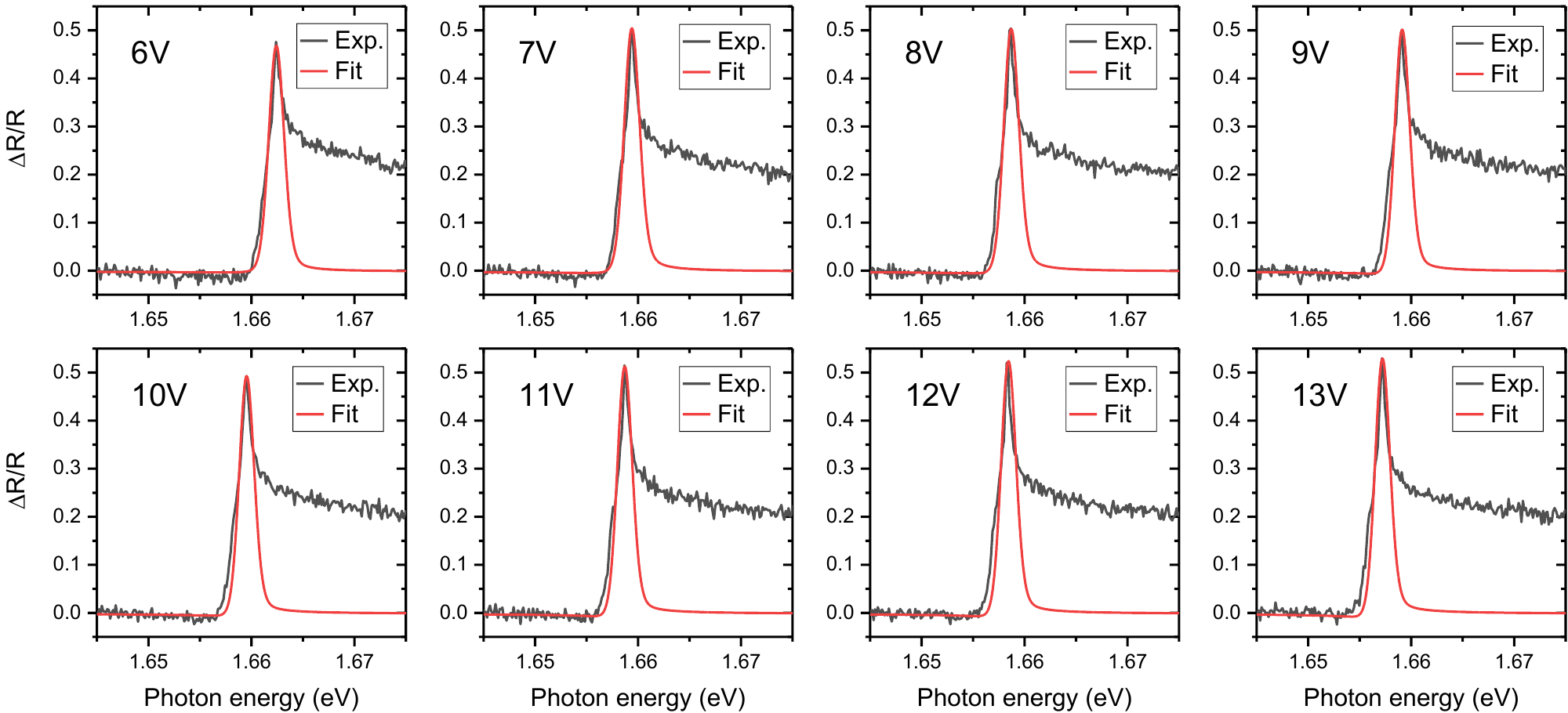}
\caption{Differential reflectivity spectra for different gate voltages}
\label{S3}
\end{figure}

\subsection{Fitting procedure of the TRPL data}

The fitting function used in Fig. 3(a) (main text) is a bi-exponential model convolved with a Gaussian, with parameters determined from a fit of the instrument response function (IRF). This approach accurately reproduces the decays across the entire time window ($-$5 to 150 ps), with the exception of the data at 5–7 V. In this lower-voltage range, the rise time is extracted from a short-time fit ($-$5 to 30 ps), while the decay time is obtained from an asymptotic fit at longer times (50 to 150 ps).\\\par

The deviations observed at low voltage may indicate the presence of additional relaxation pathways not captured by our simplified three-level model. These could be related to lower energy states (such as dark trions and their replicas) that become prominent at small voltages (see Figure 2a of the main text). However, since the primary focus of our study is on the $O$ and $M$ regimes at higher voltages—where the fits are accurate across the full time range—this limitation does not impact the overall conclusions of the work.\\\par

\subsection{Characteristic times of processes}

\begin{figure} [!h]
    \centering
    \includegraphics[width=\linewidth]{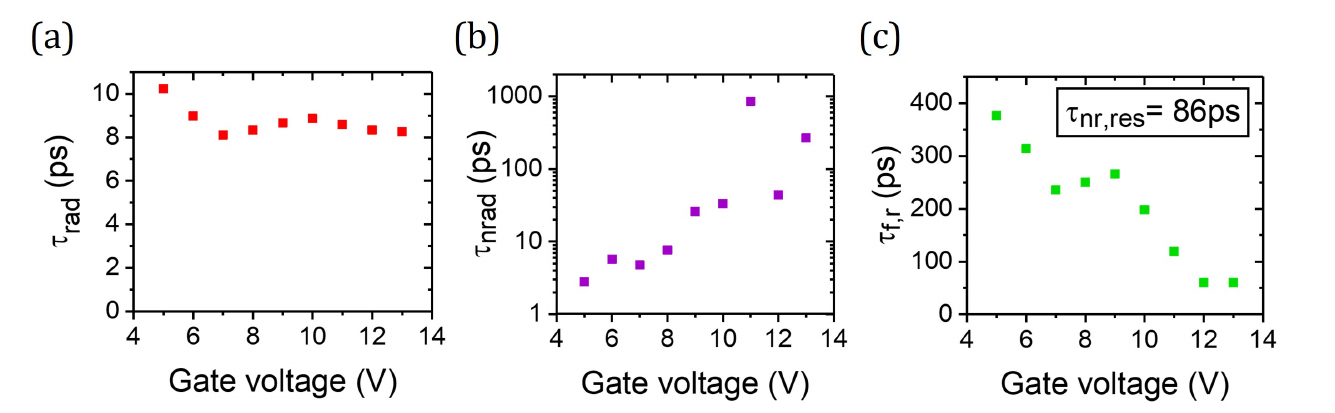}
    \caption{(a) $\tau_{\mathrm{rad}}$, (b) $\tau_{\mathrm{nrad}}$, and  (c) $\tau_{\mathrm{f,r}}$ as a function of the gate voltage.}
    \label{fig:S4}
\end{figure}

Figure~\ref{fig:S4}a,b presents the radiative and non-radiative lifetimes extracted from TRPL and reflectivity measurements. 
The radiative lifetime $\tau_\mathrm{rad}$ remains essentially constant, between 8 and 10~ps, across the $H$, $O$, and $M$ regimes. 
In contrast, the non-radiative lifetime $\tau_\mathrm{nrad}$ is shorter than $\tau_\mathrm{rad}$ in the $H$ regime, which may indicate relaxation into lower-energy states. 
Upon entering the $O$ regime, $\tau_\mathrm{nrad}$ increases with doping. 
At high doping levels it reaches very large values, consistent with the internal quantum efficiency $\eta_0$ approaching unity, although these points carry larger experimental uncertainties.\\\par

Figure~\ref{fig:S4}c shows the formation time of the cold complexes, $\tau_{\mathrm{f,r}}$, as a function of gate voltage. 
The formation accelerates as doping increases, exhibiting a pronounced plateau within the $O$ regime. 
The microscopic origin of this behavior remains unclear and will require a dedicated theoretical framework to fully elucidate the formation dynamics of the excitonic complexes.\\\par

\bibliographystyleB{unsrt}
\bibliographyB{}


\begin{thebibliography}{10}

\bibitem{Mak2010}
Kin~Fai Mak, Changgu Lee, James Hone, Jie Shan, and Tony~F. Heinz.
\newblock Atomically thin {MoS$_2$}: A new direct-gap semiconductor.
\newblock {\em Phys. Rev. Lett.}, 105:136805, Sep 2010.

\bibitem{splendiani_emerging_2010}
Andrea Splendiani, Liang Sun, Yuanbo Zhang, Tianshu Li, Jonghwan Kim, Chi-Yung Chim, Giulia Galli, and Feng Wang.
\newblock Emerging photoluminescence in monolayer {MoS$_2$}.
\newblock {\em Nano Letters}, 10(4):1271--1275, April 2010.

\bibitem{wang_colloquium_2018}
Gang Wang, Alexey Chernikov, Mikhail~M. Glazov, Tony~F. Heinz, Xavier Marie, Thierry Amand, and Bernhard Urbaszek.
\newblock Colloquium: {Excitons} in atomically thin transition metal dichalcogenides.
\newblock {\em Rev. Mod. Phys.}, 90(2):021001, April 2018.

\bibitem{goryca_revealing_2019}
M.~Goryca, J.~Li, A.~V. Stier, T.~Taniguchi, K.~Watanabe, E.~Courtade, S.~Shree, C.~Robert, B.~Urbaszek, X.~Marie, and S.~A. Crooker.
\newblock Revealing exciton masses and dielectric properties of monolayer semiconductors with high magnetic fields.
\newblock {\em Nat Commun}, 10(1):4172, September 2019.

\bibitem{chernikov_exciton_2014}
Alexey Chernikov, Timothy~C. Berkelbach, Heather~M. Hill, Albert Rigosi, Yilei Li, Burak Aslan, David~R. Reichman, Mark~S. Hybertsen, and Tony~F. Heinz.
\newblock Exciton binding energy and nonhydrogenic rydberg series in monolayer {WS$_2$}.
\newblock {\em Phys. Rev. Lett.}, 113(7):076802, August 2014.

\bibitem{qiu_optical_2013}
Diana~Y. Qiu, Felipe~H. da~Jornada, and Steven~G. Louie.
\newblock Optical spectrum of {MoS$_2$:} {Many}-body effects and diversity of exciton states.
\newblock {\em Physical Review Letters}, 111(21):216805, November 2013.

\bibitem{mak_tightly_2013}
Kin~Fai Mak, Keliang He, Changgu Lee, Gwan~Hyoung Lee, James Hone, Tony~F. Heinz, and Jie Shan.
\newblock Tightly bound trions in monolayer {MoS$_2$}.
\newblock {\em Nature Mater}, 12(3):207--211, March 2013.

\bibitem{ross_electrical_2013}
Jason~S. Ross, Sanfeng Wu, Hongyi Yu, Nirmal~J. Ghimire, Aaron~M. Jones, Grant Aivazian, Jiaqiang Yan, David~G. Mandrus, Di~Xiao, Wang Yao, and Xiaodong Xu.
\newblock Electrical control of neutral and charged excitons in a monolayer semiconductor.
\newblock {\em Nat Commun}, 4(1):1474, February 2013.

\bibitem{courtade_charged_2017}
E.~Courtade, M.~Semina, M.~Manca, M.~M. Glazov, C.~Robert, F.~Cadiz, G.~Wang, T.~Taniguchi, K.~Watanabe, M.~Pierre, W.~Escoffier, E.~L. Ivchenko, P.~Renucci, X.~Marie, T.~Amand, and B.~Urbaszek.
\newblock Charged excitons in monolayer {WSe$_2$}: {Experiment} and theory.
\newblock {\em Phys. Rev. B}, 96(8):085302, August 2017.

\bibitem{perea-causin_trion_2024}
Raul Perea-Causin, Samuel Brem, Ole Schmidt, and Ermin Malic.
\newblock Trion photoluminescence and trion stability in atomically thin semiconductors.
\newblock {\em Physical Review Letters}, 132(3):036903, January 2024.

\bibitem{liu_gate_2019}
Erfu Liu, Jeremiah van Baren, Zhengguang Lu, Mashael~M. Altaiary, Takashi Taniguchi, Kenji Watanabe, Dmitry Smirnov, and Chun~Hung Lui.
\newblock Gate tunable dark trions in monolayer {WSe$_2$}.
\newblock {\em Phys. Rev. Lett.}, 123(2):027401, July 2019.

\bibitem{li_direct_2019}
Zhipeng Li, Tianmeng Wang, Zhengguang Lu, Mandeep Khatoniar, Zhen Lian, Yuze Meng, Mark Blei, Takashi Taniguchi, Kenji Watanabe, Stephen~A. McGill, Sefaattin Tongay, Vinod~M. Menon, Dmitry Smirnov, and Su-Fei Shi.
\newblock Direct observation of gate-tunable dark trions in monolayer {WSe$_2$}.
\newblock {\em Nano Lett.}, 19(10):6886--6893, October 2019.

\bibitem{yang_relaxation_2022}
Min Yang, Lei Ren, Cedric Robert, Dinh Van~Tuan, Laurent Lombez, Bernhard Urbaszek, Xavier Marie, and Hanan Dery.
\newblock Relaxation and darkening of excitonic complexes in electrostatically doped monolayer {WSe$_2$}: {Roles} of exciton-electron and trion-electron interactions.
\newblock {\em Phys. Rev. B}, 105(8):085302, February 2022.

\bibitem{malic_dark_2018}
Ermin Malic, Malte Selig, Maja Feierabend, Samuel Brem, Dominik Christiansen, Florian Wendler, Andreas Knorr, and Gunnar BerghÃ¤user.
\newblock Dark excitons in transition metal dichalcogenides.
\newblock {\em Physical Review Materials}, 2(1):014002, January 2018.

\bibitem{ye_efficient_2018}
Ziliang Ye, Lutz Waldecker, Eric~Yue Ma, Daniel Rhodes, Abhinandan Antony, Bumho Kim, Xiao-Xiao Zhang, Minda Deng, Yuxuan Jiang, Zhengguang Lu, Dmitry Smirnov, Kenji Watanabe, Takashi Taniguchi, James Hone, and Tony~F. Heinz.
\newblock Efficient generation of neutral and charged biexcitons in encapsulated {WSe$_2$} monolayers.
\newblock {\em Nature Communications}, 9(1):3718, September 2018.

\bibitem{molas_brightening_2017}
M~R Molas, C~Faugeras, A~O Slobodeniuk, K~Nogajewski, M~Bartos, D~M Basko, and M~Potemski.
\newblock Brightening of dark excitons in monolayers of semiconducting transition metal dichalcogenides.
\newblock {\em 2D Materials}, 4(2):021003, January 2017.

\bibitem{liu_two-step_2023}
Song Liu, Yang Liu, Luke Holtzman, Baichang Li, Madisen Holbrook, Jordan Pack, Takashi Taniguchi, Kenji Watanabe, Cory~R. Dean, Abhay~N. Pasupathy, Katayun Barmak, Daniel~A. Rhodes, and James Hone.
\newblock Two-step flux synthesis of ultrapure transition-metal dichalcogenides.
\newblock {\em ACS Nano}, 17(17):16587--16596, September 2023.

\bibitem{goodman_substrate-dependent_2020}
A.~J. Goodman, D.-H. Lien, G.~H. Ahn, L.~L. Spiegel, M.~Amani, A.~P. Willard, A.~Javey, and W.~A. Tisdale.
\newblock Substrate-dependent exciton diffusion and annihilation in chemically treated {MoS$_2$} and {WS$_2$}.
\newblock {\em J. Phys. Chem. C}, 124(22):12175--12184, June 2020.

\bibitem{amani_near-unity_2015}
Matin Amani, Der-Hsien Lien, Daisuke Kiriya, Jun Xiao, Angelica Azcatl, Jiyoung Noh, Surabhi~R. Madhvapathy, Rafik Addou, Santosh KC, Madan Dubey, Kyeongjae Cho, Robert~M. Wallace, Si-Chen Lee, Jr-Hau He, Joel~W. Ager, Xiang Zhang, Eli Yablonovitch, and Ali Javey.
\newblock Near-unity photoluminescence quantum yield in {MoS$_2$}.
\newblock {\em Science}, 350(6264):1065--1068, November 2015.

\bibitem{tanoh_giant_2021}
Arelo O.~A. Tanoh, Jack Alexander-Webber, Ye~Fan, Nicholas Gauriot, James Xiao, Raj Pandya, Zhaojun Li, Stephan Hofmann, and Akshay Rao.
\newblock Giant photoluminescence enhancement in {MoSe$_2$} monolayers treated with oleic acid ligands.
\newblock {\em Nanoscale Advances}, 3(14):4216--4225, July 2021.

\bibitem{lien_electrical_2019}
Der-Hsien Lien, Shiekh~Zia Uddin, Matthew Yeh, Matin Amani, Hyungjin Kim, Joel~W. Ager, Eli Yablonovitch, and Ali Javey.
\newblock Electrical suppression of all nonradiative recombination pathways in monolayer semiconductors.
\newblock {\em Science}, 364(6439):468--471, May 2019.

\bibitem{van_tuan_six-body_2022}
Dinh Van~Tuan, Su-Fei Shi, Xiaodong Xu, Scott~A. Crooker, and Hanan Dery.
\newblock Six-body and eight-body exciton states in monolayer {WSe$_2$}.
\newblock {\em Phys. Rev. Lett.}, 129(7):076801, August 2022.

\bibitem{choi_emergence_2024}
J.~Choi, J.~Li, D.~Van~Tuan, H.~Dery, and S.~A. Crooker.
\newblock Emergence of composite many-body exciton states in {WS$_2$} and {MoSe$_2$} monolayers.
\newblock {\em Physical Review B}, 109(4):L041304, January 2024.

\bibitem{dijkstra_ten-valley_2025}
Alain Dijkstra, Amine~Ben Mhenni, Dinh~Van Tuan, Elif Çetiner, Muriel Schur-Wilkens, Junghwan Kim, Laurin Steiner, Kenji Watanabe, Takashi Taniguchi, Matteo Barbone, Nathan~P. Wilson, Hanan Dery, and Jonathan~J. Finley.
\newblock Ten-valley excitonic complexes in charge-tunable monolayer {WSe$_2$}, May 2025.
\newblock arXiv:2505.08923 [cond-mat].

\bibitem{pierret_dielectric_2022}
A~Pierret, D~Mele, H~Graef, J~Palomo, T~Taniguchi, K~Watanabe, Y~Li, B~Toury, C~Journet, P~Steyer, V~Garnier, A~Loiseau, J-M Berroir, E~Bocquillon, G~Fève, C~Voisin, E~Baudin, M~Rosticher, and B~Plaçais.
\newblock Dielectric permittivity, conductivity and breakdown field of hexagonal boron nitride.
\newblock {\em Mater. Res. Express}, 9(6):065901, June 2022.

\bibitem{kapuscinski_rydberg_2021}
Piotr Kapuściński, Alex Delhomme, Diana Vaclavkova, Artur~O. Slobodeniuk, Magdalena Grzeszczyk, Miroslav Bartos, Kenji Watanabe, Takashi Taniguchi, Clément Faugeras, and Marek Potemski.
\newblock Rydberg series of dark excitons and the conduction band spin-orbit splitting in monolayer {WSe$_2$}.
\newblock {\em Commun Phys}, 4(1):186, August 2021.

\bibitem{stier_magnetooptics_2018}
Andreas~V Stier, Nathan~P Wilson, Kirill~A Velizhanin, Junichiro Kono, Xiaodong Xu, and Scott~A Crooker.
\newblock Magnetooptics of exciton {Rydberg} states in a monolayer semiconductor.
\newblock {\em Physical Review Letters}, 120(5):057405, 2018.

\bibitem{efimkin_many-body_2017}
Dmitry~K. Efimkin and Allan~H. MacDonald.
\newblock Many-body theory of trion absorption features in two-dimensional semiconductors.
\newblock {\em Phys. Rev. B}, 95(3):035417, January 2017.

\bibitem{sidler_fermi_2017}
Meinrad Sidler, Patrick Back, Ovidiu Cotlet, Ajit Srivastava, Thomas Fink, Martin Kroner, Eugene Demler, and Atac Imamoglu.
\newblock Fermi polaron-polaritons in charge-tunable atomically thin semiconductors.
\newblock {\em Nature Phys}, 13(3):255--261, March 2017.

\bibitem{glazov_optical_2020}
M.~M. Glazov.
\newblock Optical properties of charged excitons in two-dimensional semiconductors.
\newblock {\em J. Chem. Phys.}, 153(3):034703, July 2020.

\bibitem{suris_excitons_2001}
R.a. Suris, V.p. Kochereshko, G.v. Astakhov, D.r. Yakovlev, W.~Ossau, J.~Nürnberger, W.~Faschinger, G.~Landwehr, T.~Wojtowicz, G.~Karczewski, and J.~Kossut.
\newblock Excitons and trions modified by interaction with a two-dimensional electron gas.
\newblock {\em physica status solidi (b)}, 227(2):343--352, 2001.

\bibitem{bronold_absorption_2000}
F.X. Bronold.
\newblock Absorption spectrum of a weakly n-doped semiconductor quantum well.
\newblock {\em Phys. Rev. B}, 61(19):12620--12623, May 2000.

\bibitem{chang_crossover_2018}
Yia-Chung Chang, Shiue-Yuan Shiau, and Monique Combescot.
\newblock Crossover from trion-hole complex to exciton-polaron in n-doped two-dimensional semiconductor quantum wells.
\newblock {\em Phys. Rev. B}, 98(23):235203, December 2018.

\bibitem{jones_optical_2013}
Aaron~M. Jones, Hongyi Yu, Nirmal~J. Ghimire, Sanfeng Wu, Grant Aivazian, Jason~S. Ross, Bo~Zhao, Jiaqiang Yan, David~G. Mandrus, Di~Xiao, Wang Yao, and Xiaodong Xu.
\newblock Optical generation of excitonic valley coherence in monolayer {WSe$_2$}.
\newblock {\em Nature Nanotechnology}, 8(9):634--638, September 2013.

\bibitem{wang_probing_2017}
Zefang Wang, Liang Zhao, Kin~Fai Mak, and Jie Shan.
\newblock Probing the spin-polarized electronic band structure in monolayer transition metal dichalcogenides by optical spectroscopy.
\newblock {\em Nano letters}, 17(2):740--746, 2017.

\bibitem{li_many-body_2022}
Jing Li, Mateusz Goryca, Junho Choi, Xiaodong Xu, and Scott~A. Crooker.
\newblock Many-body exciton and intervalley correlations in heavily electron-doped {WSe$_2$} monolayers.
\newblock {\em Nano Letters}, 22(1):426--432, January 2022.

\bibitem{ren_measurement_2023}
Lei Ren, Cedric Robert, Hanan Dery, Minhao He, Pengke Li, Dinh Van~Tuan, Pierre Renucci, Delphine Lagarde, Takashi Taniguchi, Kenji Watanabe, Xiaodong Xu, and Xavier Marie.
\newblock Measurement of the conduction band spin-orbit splitting in {WSe$_2$} and {WS$_2$} monolayers.
\newblock {\em Phys. Rev. B}, 107(24):245407, June 2023.

\bibitem{jindal_brightened_2025}
V.~Jindal, K.~Mourzidis, A.~Balocchi, C.~Robert, P.~Li, D.~Van~Tuan, L.~Lombez, D.~Lagarde, P.~Renucci, T.~Taniguchi, K.~Watanabe, H.~Dery, and X.~Marie.
\newblock Brightened emission of dark trions in transition metal dichalcogenide monolayers.
\newblock {\em Phys. Rev. B}, 111(15):155409, April 2025.

\bibitem{dery_energy_2025}
Hanan Dery, Cedric Robert, Scott~A. Crooker, Xavier Marie, and Dinh Van~Tuan.
\newblock Energy shifts and broadening of excitonic resonances in electrostatically doped semiconductors.
\newblock {\em Phys. Rev. X}, 15(3):031049, August 2025.

\bibitem{kormanyos_kp_2015}
Andor Kormányos, Guido Burkard, Martin Gmitra, Jaroslav Fabian, Viktor Zólyomi, Neil~D Drummond, and Vladimir Fal’ko.
\newblock k·p theory for two-dimensional transition metal dichalcogenide semiconductors.
\newblock {\em 2D Mater.}, 2(2):022001, April 2015.

\bibitem{fang_control_2019}
H.~H. Fang, B.~Han, C.~Robert, M.~A. Semina, D.~Lagarde, E.~Courtade, T.~Taniguchi, K.~Watanabe, T.~Amand, B.~Urbaszek, M.~M. Glazov, and X.~Marie.
\newblock Control of the exciton radiative lifetime in van der {Waals} heterostructures.
\newblock {\em Phys. Rev. Lett.}, 123(6):067401, August 2019.

\bibitem{robert_optical_2018}
C.~Robert, M.~A. Semina, F.~Cadiz, M.~Manca, E.~Courtade, T.~Taniguchi, K.~Watanabe, H.~Cai, S.~Tongay, B.~Lassagne, P.~Renucci, T.~Amand, X.~Marie, M.~M. Glazov, and B.~Urbaszek.
\newblock Optical spectroscopy of excited exciton states in {MoS$_2$} monolayers in van der {Waals} heterostructures.
\newblock {\em Physical Review Materials}, 2(1):011001, January 2018.

\end{thebibliography}


\begin{thebibliography}{1}

\bibitem{taniguchi2007synthesis}
Takashi Taniguchi and Kenji Watanabe.
\newblock Synthesis of high-purity boron nitride single crystals under high pressure by using ba--bn solvent.
\newblock {\em Journal of crystal growth}, 303(2):525--529, 2007.

\bibitem{castellanos2014deterministic}
Andres Castellanos-Gomez, Michele Buscema, Rianda Molenaar, Vibhor Singh, Laurens Janssen, Herre~SJ Van Der~Zant, and Gary~A Steele.
\newblock Deterministic transfer of two-dimensional materials by all-dry viscoelastic stamping.
\newblock {\em 2D Materials}, 1(1):011002, 2014.

\bibitem{robert_optical_2018}
C.~Robert, M.~A. Semina, F.~Cadiz, M.~Manca, E.~Courtade, T.~Taniguchi, K.~Watanabe, H.~Cai, S.~Tongay, B.~Lassagne, P.~Renucci, T.~Amand, X.~Marie, M.~M. Glazov, and B.~Urbaszek.
\newblock Optical spectroscopy of excited exciton states in {MoS2} monolayers in van der {Waals} heterostructures.
\newblock {\em Physical Review Materials}, 2(1):011001, January 2018.

\end{thebibliography}


\begin{thebibliography}{0}%
\makeatletter
\providecommand \@ifxundefined [1]{%
 \@ifx{#1\undefined}
}%
\providecommand \@ifnum [1]{%
 \ifnum #1\expandafter \@firstoftwo
 \else \expandafter \@secondoftwo
 \fi
}%
\providecommand \@ifx [1]{%
 \ifx #1\expandafter \@firstoftwo
 \else \expandafter \@secondoftwo
 \fi
}%
\providecommand \natexlab [1]{#1}%
\providecommand \enquote  [1]{``#1''}%
\providecommand \bibnamefont  [1]{#1}%
\providecommand \bibfnamefont [1]{#1}%
\providecommand \citenamefont [1]{#1}%
\providecommand \href@noop [0]{\@secondoftwo}%
\providecommand \href [0]{\begingroup \@sanitize@url \@href}%
\providecommand \@href[1]{\@@startlink{#1}\@@href}%
\providecommand \@@href[1]{\endgroup#1\@@endlink}%
\providecommand \@sanitize@url [0]{\catcode `\\12\catcode `\$12\catcode `\&12\catcode `\#12\catcode `\^12\catcode `\_12\catcode `\%12\relax}%
\providecommand \@@startlink[1]{}%
\providecommand \@@endlink[0]{}%
\providecommand \url  [0]{\begingroup\@sanitize@url \@url }%
\providecommand \@url [1]{\endgroup\@href {#1}{\urlprefix }}%
\providecommand \urlprefix  [0]{URL }%
\providecommand \Eprint [0]{\href }%
\providecommand \doibase [0]{https://doi.org/}%
\providecommand \selectlanguage [0]{\@gobble}%
\providecommand \bibinfo  [0]{\@secondoftwo}%
\providecommand \bibfield  [0]{\@secondoftwo}%
\providecommand \translation [1]{[#1]}%
\providecommand \BibitemOpen [0]{}%
\providecommand \bibitemStop [0]{}%
\providecommand \bibitemNoStop [0]{.\EOS\space}%
\providecommand \EOS [0]{\spacefactor3000\relax}%
\providecommand \BibitemShut  [1]{\csname bibitem#1\endcsname}%
\let\auto@bib@innerbib\@empty
\end{thebibliography}%
\end{document}